\documentclass[preprint,aps,showpacs,showkeys]{revtex4}

\usepackage{amsfonts}
\usepackage{amssymb,amsmath}
\usepackage{graphicx}
\usepackage{dcolumn}
\usepackage{bm}
\usepackage{color}

\renewcommand{\theequation}{\arabic{section}.\arabic{equation}}

\begin{document}

\title{The square-kagome quantum Heisenberg antiferromagnet at high magnetic fields:
       The localized-magnon paradigm and beyond}

\author{Oleg Derzhko}
\affiliation{Institute for Condensed Matter Physics,
          National Academy of Sciences of Ukraine,
          1 Svientsitskii Street, L'viv-11, 79011, Ukraine}
\affiliation{Department for Theoretical Physics, 
          Ivan Franko National University of L'viv, 
          12 Drahomanov Street, L'viv-5, 79005, Ukraine}
\affiliation{Abdus Salam International Centre for Theoretical Physics,
          Strada Costiera 11, I-34151 Trieste, Italy}

\author{Johannes Richter}
\affiliation{Institut f\"{u}r theoretische Physik,
          Otto-von-Guericke-Universit\"{a}t Magdeburg,
          P.O. Box 4120, D-39016 Magdeburg, Germany}

\author{Olesia Krupnitska}
\affiliation{Institute for Condensed Matter Physics,
          National Academy of Sciences of Ukraine,
          1 Svientsitskii Street, L'viv-11, 79011, Ukraine}

\author{Taras Krokhmalskii}
\affiliation{Institute for Condensed Matter Physics,
          National Academy of Sciences of Ukraine,
          1 Svientsitskii Street, L'viv-11, 79011, Ukraine}
\affiliation{Department for Theoretical Physics, 
          Ivan Franko National University of L'viv, 
          12 Drahomanov Street, L'viv-5, 79005, Ukraine}

\date{\today}

\pacs{75.10.Jm}

\keywords{quantum Heisenberg antiferromagnet,
          square-kagome lattice,
          localized magnons,
          Berezinskii-Kosterlitz-Thouless transition}

\begin{abstract}
We consider the spin-1/2 antiferromagnetic Heisenberg model on the two-dimensional square-kagome lattice 
with almost dispersionless lowest magnon band.
For a general exchange coupling geometry 
we elaborate low-energy effective Hamiltonians which emerge at high magnetic fields.
The effective model to describe the low-energy degrees of freedom of the initial frustrated quantum spin model  
is the (unfrustrated) square-lattice spin-1/2 $XXZ$ model in a $z$-aligned magnetic field. 
For the effective model we perform quantum Monte Carlo simulations
to discuss the low-temperature properties of the square-kagome quantum Heisenberg antiferromagnet at high magnetic fields.
We pay special attention to a magnetic-field driven Berezinskii-Kosterlitz-Thouless phase transition which occurs at low temperatures.
\end{abstract}

\maketitle

\section{Introduction}
\label{sec1}
\setcounter{equation}{0}

Antiferromagnetic interactions between the spins carried by magnetic ions placed on a nonbipartite lattice 
(like the triangle lattice or the kagome lattice) 
are competing, i.e., frustrated.  
The Zeeman interaction with an external magnetic field introduces even more competition.  
As a result, 
the quantum Heisenberg antiferromagnet on a low-dimensional nonbipartite lattice in a magnetic field 
provides an excellent playground for the study of the interplay between quantum fluctuations and frustration.  
In such systems new phenomena may emerge.  
Therefore, the study of frustrated quantum antiferromagnets attracts much attention nowadays.\cite{lnp} 
Interestingly, 
in some cases frustrated quantum Heisenberg antiferromagnets 
admit a rather detailed study of their low-temperature properties at high magnetic
fields,
namely for the so-called localized-magnon systems which have a dispersionless (flat) lowest magnon band.\cite{lm,SP} 
It has been shown that the localized-magnon systems in the high-field low-temperature regime 
may be understood 
using specific methods of classical statistical mechanics.\cite{lm,localized_magnons1,SP,localized_magnons2,epjb2006,bilayer,double_tetrahedra_epjb}
However, this classical description of the localized-magnon quantum spin systems 
was developed under the assumption of the so-called ideal geometry,
i.e., the conditions for localization of the magnon states are strictly fulfilled 
(i.e., the lowest magnon band is strictly flat).  
As a rule, this assumption is violated in real-life systems.  
Hence, the case of nonideal geometry,  
when the localization condition is (slightly) violated,
is more relevant from the experimental point of view. 
There were several papers related to this nonideal situation,\cite{distorted_ladder,frbi3,effective_xy1,effective_xy2,drk} 
which, however, did not use the localized-magnon paradigm.

\begin{figure}[ht]
\begin{center}
\includegraphics[clip=on,width=85mm,angle=0]{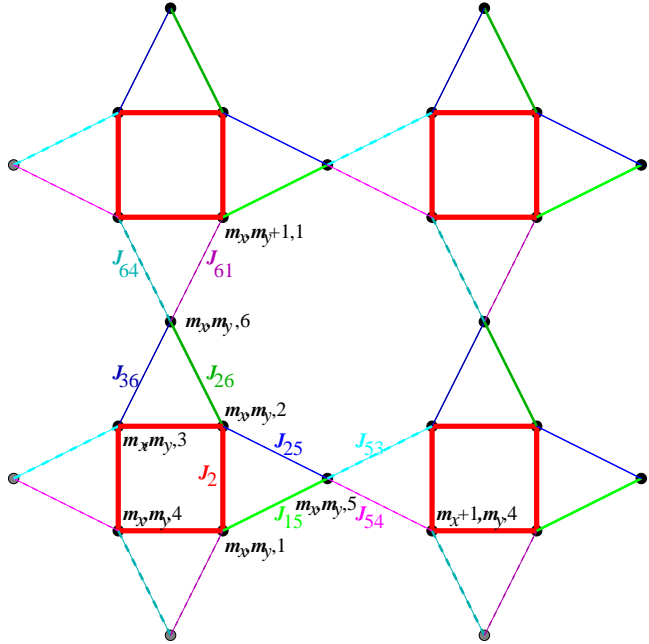}
\caption
{(Color online) 
The square-kagome lattice described by Hamiltonian (\ref{201}).
The trapping cells for localized magnons (squares) are indicated by bold red lines ($J_2$ bonds).}
\label{fig01}
\end{center}
\end{figure}

Recently\cite{drkk} we have developed a systematic treatment of a certain class of localized-magnon systems, 
namely the monomer class,\cite{epjb2006}
to consider  small deviations from ideal geometry. 
In particular,
we have investigated the antiferromagnetic quantum Heisenberg model in a field 
on the diamond chain, the dimer-plaquette chain, and the square-kagome lattice 
(for the latter lattice see Fig.~\ref{fig01}). 
We mention that all of these models of frustrated quantum antiferromagnets 
have attracted attention previously as strongly frustrated quantum spin systems, 
and they were investigated by various authors also at zero or moderate fields, where the localized-magnon scenario is not relevant, 
see Refs.~\onlinecite{diamond,dim_pla,sqkag}.  
Inspired by the situation in the diamond-chain like compound azurite,\cite{Kikuchi,azurite-parameters,effective_xy2}
in Ref.~\onlinecite{drkk} it was assumed that 
$J_{25}=J_{54}=J_{36}=J_{61}=J_1$
and 
$J_{15}=J_{53}=J_{26}=J_{64}=J_3$ (cf. Fig.~\ref{fig01})
that we will call azurite-like geometry.
For that type of exchange geometry, by eliminating high-energy degrees of freedom, 
we constructed several low-energy effective Hamiltonians which are much simpler to treat than the initial ones.
Thus for the $N$-site frustrated square-kagome lattice 
with the azurite-like deviation from ideal geometry\cite{azurite-parameters,effective_xy2} 
we obtained the Hamiltonian of the ${\cal{N}}$-site (${\cal{N}}=N/6$) unfrustrated square-lattice (pseudo)spin-1/2 $XXZ$ model 
in a $z$-aligned magnetic field.
Then we performed exact-diagonalization studies for the obtained effective model of ${\cal{N}}=20$ sites
(corresponding to $N=120$ sites for the initial square-kagome system)
to discuss the low-temperature properties of the spin-1/2 square-kagome Heisenberg antiferromagnet in a field.
The most intriguing feature that we found in Ref.~\onlinecite{drkk}
is the existence  of a magnetic-field driven Berezinskii-Kosterlitz-Thouless (BKT) phase transition at low temperatures.

The aim of the present paper, 
which continues the preceeding study\cite{drkk} with a special focus on the square-kagome system,
is three-fold.
First,
we will provide effective Hamiltonians for a general exchange coupling scheme, 
see Fig.~\ref{fig01}, 
going beyond the the azurite-like geometry.
(Besides, we will report in Appendix~\ref{a} similar results for the one-dimensional diamond-chain case.)
Second,
instead of the exact-diagonalization method that is restricted to small systems only,
we now present data obtained by quantum Monte Carlo simulations 
for the effective model of much larger size up to ${\cal{N}}=24\times24=576$ sites
(corresponding to $N=3456$ sites for the initial square-kagome system).
Based on  these data 
we are able to make more precise predictions for the high-field low-temperature properties of the square-kagome quantum Heisenberg antiferromagnet,
in particular, for the phase diagram of the model.
Third,
we will provide a more detailed discussion of the BKT phase transition
which may occur in the square-kagome quantum Heisenberg antiferromagnet 
emphasizing some tasks for further studies.

\section{Low-energy effective Hamiltonians at high magnetic fields}
\label{sec2}
\setcounter{equation}{0}

In this paper,
we consider the standard spin-1/2 antiferromagnetic isotropic Heisenberg model in a magnetic field 
with the Hamiltonian
\begin{eqnarray}
\label{201}
H=\sum_{(ij)} J_{ij} {\bf{s}}_i \cdot {\bf{s}}_j-hS^z,
\;\;
S^z=\sum_{i=1}^Ns_i^z, 
\;\; 
J_{ij} > 0.
\end{eqnarray}
Here the first sum runs over all nearest-neighbor bonds on the square-kagome lattice,
whereas the second one runs over all $N$ lattice sites.
It is convenient to label the sites by a pair of indeces, 
where the first vector index ${\bf{m}}=(m_x,m_y)$ enumerates the ${\cal{N}}=N/6$ unit cells
and the second one enumerates the position of the site within the unit cell, 
see Fig.~\ref{fig01}.
Since $[S^z,H]=0$, the eigenvalues of $S^z$ are good quantum numbers.
We consider magnetic fields in the vicinity of the saturation field $h_{\rm{sat}}$. 
For the ideal geometry, when $J_{15}=\ldots=J_{64}=J\le J_2$, we have
$h_{\rm{sat}}=h_1=2J_2+J$. Then for $h>h_1$ the ground state is the fully polarized ferromagnetic state, 
and the band of the lowest-magnon excitations is dispersionless (flat).
An eigenstate from this band can be written as a localized-magnon state,\cite{lm}
where the spin-flip (magnon) is trapped on a square (trapping cell), 
see Fig.~\ref{fig01}.
Owing to the localized nature of these states 
the many-magnon states in the subspaces $S^z=N/2-2,\ldots,N/2-{\cal{N}}$ can be constructed by filling the traps by localized magnons. 
Clearly, all these states are linear independent.\cite{linear_in}
Moreover, these localized-magnon states have the lowest energy in their corresponding $S^z$-subspace,  
if the strength of the antiferromagnetic bonds of the trapping cells $J_2$ exceeds a lower bound.\cite{lm,Schmidt}
The degeneracy of the localized-magnon states is calculated 
via mapping of these states onto spatial configurations of hard monomers on an auxiliary square lattice.\cite{localized_magnons2,epjb2006}
At low temperatures and for magnetic fields $h$ around the saturation field $h_{\rm{sat}}=h_1$
the contribution of localized-magnon states dominates the partition function.\cite{epjb2006}

In what follows we consider the most general violation of the ideal geometry 
by allowing that all values of $J_{15},\ldots,J_{64}$ are  different,
see Fig.~\ref{fig01}.
However, we  assume that the deviations from ideal geometry are not too large,
i.e., perturbation theory is applicable.
To derive the effective Hamiltonian we follow closely the lines given in Ref.~\onlinecite{drkk}.

At high fields considered here, only a few states of the trapping cell are relevant,
namely,
the fully polarized state 
$\vert u\rangle= \vert \uparrow_1\uparrow_2\uparrow_3\uparrow_4\rangle$ with the energy $J_2-2h$
and the one-magnon state 
$\vert d\rangle=\left(\vert \uparrow_1\uparrow_2\uparrow_3\downarrow_4\rangle 
-\vert \uparrow_1\uparrow_2\downarrow_3\uparrow_4\rangle 
+\vert \uparrow_1\downarrow_2\uparrow_3\uparrow_4\rangle 
-\vert \downarrow_1\uparrow_2\uparrow_3\uparrow_4\rangle\right)/2$ with the energy $-J_2-h$.
All other sites carry fully polarized (i.e., $z$-aligned) spins.
Decreasing the magnetic field from $h>h_1$ to $h<h_1$, 
in the case of ideal geometry,
the ground state of the cell undergoes a transition 
from the state $\vert u\rangle$ to the state $\vert d\rangle$ 
at the saturation field $h_{{\rm{sat}}}=h_1$.
Therefore it is a reasonable approximation to take into account further  
only these 2 most relevant states $\vert u\rangle$ and $\vert d\rangle$ for each square 
instead of the complete set of 16 states of a square.
According to Ref.~\onlinecite{drkk}, we use this restricted set of states
and consider as the starting point instead of $H$ (\ref{201}) the projected Hamiltonian
\begin{eqnarray}
\label{202}
{\cal{H}}={\cal{P}}H{\cal{P}},
\nonumber\\
{\cal{P}}=\otimes_{{\bf{m}}}{\cal{P}}_{{\bf{m}}},
\;\;\;
{\cal{P}}_{{\bf{m}}}
=\left(\vert u\rangle\langle u\vert+\vert d\rangle\langle d\vert\right)_{\bf{m}}.
\end{eqnarray}
Here ${\cal{P}}_{\bf{m}}$ is the projector on the relevant states of the trapping cell ${\bf{m}}$.
Introducing (pseudo)spin-1/2 operators for each cell,
\begin{eqnarray}
\label{203}
T^z=\frac{1}{2}\left(\vert u\rangle\langle u\vert-\vert d\rangle\langle d\vert\right),
T^+=\vert u\rangle\langle d\vert,
T^-=\vert d\rangle\langle u\vert,
\end{eqnarray}
we can write the Hamiltonian ${\cal{H}}$ in Eq.~(\ref{202}) as
\begin{eqnarray}
\label{204}
{\cal{H}}=\sum_{{\bf{m}}}
\left[-\frac{3}{2}h
-\left(h-2J_2\right)T^z_{\bf{m}}
\right.
\nonumber\\
\left.
-\left(h-\frac{3}{2}J_h\right)s_{{\bf{m}},5}^z
-\left(h-\frac{3}{2}J_v\right)s_{{\bf{m}},6}^z
\right.
\nonumber\\
\left.
+\frac{J_{15}+J_{25}}{4}T^z_{\bf{m}}s_{{\bf{m}},5}^z
\right.
\nonumber\\
\left.
+\frac{-J_{15}+J_{25}}{2}\left(T^x_{\bf{m}}s_{{\bf{m}},5}^x+T^y_{\bf{m}}s_{{\bf{m}},5}^y\right)
\right.
\nonumber\\
\left.
+\frac{J_{54}+J_{53}}{4}s_{{\bf{m}},5}^zT^z_{m_x+1,m_y}
\right.
\nonumber\\
\left.
+\frac{J_{54}-J_{53}}{2}\left(s_{{\bf{m}},5}^xT^x_{m_x+1,m_y}+s_{{\bf{m}},5}^yT^y_{m_x+1,m_y}\right)
\right.
\nonumber\\
\left.
+\frac{J_{26}+J_{36}}{4}T^z_{\bf{m}}s_{{\bf{m}},6}^z
\right.
\nonumber\\
\left.
+\frac{J_{26}-J_{36}}{2}\left(T^x_{\bf{m}}s_{{\bf{m}},6}^x+T^y_{\bf{m}}s_{{\bf{m}},6}^y\right)
\right.
\nonumber\\
\left.
+\frac{J_{61}+J_{64}}{4}s_{{\bf{m}},6}^zT^z_{m_x,m_y+1}
\right.
\nonumber\\
\left.
+\frac{-J_{61}+J_{64}}{2}\left(s_{{\bf{m}},6}^xT^x_{m_x,m_y+1}+s_{{\bf{m}},6}^yT^y_{m_x,m_y+1}\right)
\right],
\nonumber\\
J_h=\frac{J_{15}+J_{25}+J_{54}+J_{53}}{4},
\nonumber\\
J_v=\frac{J_{26}+J_{36}+J_{61}+J_{64}}{4}.
\end{eqnarray}
This Hamiltonian $\cal H$  corresponds to a spin-1/2 $XXZ$ model on a decorated square lattice 
(which is also known as the Lieb lattice\cite{lieb_lattice}).

Although the obtained effective model (\ref{204}) is unfrustrated and therefore is much easier to study
(for example, using quantum Monte Carlo simulations),
it can be further simplified by eliminating the spin variables 
${\bf s}_{{\bf  m},5}$ and ${\bf s}_{{\bf  m},6}$
belonging to the sites which connect the squares 
by treating small deviations from the ideal geometry perturbatively.
More specifically,
the Hamiltonian ${\cal{H}}$ given in Eq.~(\ref{204}) is separated into 
a ``main'' part ${\cal{H}}_{\rm{main}}$
\begin{eqnarray}
\label{205}
{\cal{H}}_{\rm{main}}
=\sum_{{\bf{m}}}
\left[-\frac{3}{2}h_1
-\left(h_1-2J_2\right)T^z_{\bf{m}}
\right.
\nonumber\\
\left.
-\left(h_1-\frac{3}{2}J\right)\left(s_{{\bf{m}},5}^z+s_{{\bf{m}},6}^z\right)
\right.
\nonumber\\
\left.
+\frac{J}{2}
\left(T^z_{\bf{m}}s_{{\bf{m}},5}^z+s_{{\bf{m}},5}^zT^z_{m_x+1,m_y}
\right.
\right.
\nonumber\\
\left.
\left.
+T^z_{\bf{m}}s_{{\bf{m}},6}^z+s_{{\bf{m}},6}^zT^z_{m_x,m_y+1}\right)
\right]
\end{eqnarray}
[i.e., the Hamiltonian ${\cal{H}}$ for the ideal geometry case $J_{15}=\ldots=J_{64}=J=(J_{15}+\ldots+J_{64})/8$ at $h=h_1=2J_2+J$]
and 
a ``perturbation'' ${\cal{V}}$
\begin{eqnarray}
\label{206}
{\cal{V}}
=\sum_{{\bf{m}}}
\left\{
-\frac{3}{2}\left(h-h_1\right)
-\left(h-h_1\right)T^z_{\bf{m}}
\right.
\nonumber\\
\left.
-\left[h-h_1-\frac{3}{2}\left(J_h-J\right)\right]s_{{\bf{m}},5}^z
\right.
\nonumber\\
\left.
-\left[h-h_1-\frac{3}{2}\left(J_v-J\right)\right]s_{{\bf{m}},6}^z
\right.
\nonumber\\
\left.
+\frac{J_{15}+J_{25}-2J}{4}T^z_{\bf{m}}s_{{\bf{m}},5}^z
\right.
\nonumber\\
\left.
+\frac{-J_{15}+J_{25}}{2}\left(T^x_{\bf{m}}s_{{\bf{m}},5}^x+T^y_{\bf{m}}s_{{\bf{m}},5}^y\right)
\right.
\nonumber\\
\left.
+\frac{J_{54}+J_{53}-2J}{4}s_{{\bf{m}},5}^zT^z_{m_x+1,m_y}
\right.
\nonumber\\
\left.
+\frac{J_{54}-J_{53}}{2}\left(s_{{\bf{m}},5}^xT^x_{m_x+1,m_y}+s_{{\bf{m}},5}^yT^y_{m_x+1,m_y}\right)
\right.
\nonumber\\
\left.
+\frac{J_{26}+J_{36}-2J}{4}T^z_{\bf{m}}s_{{\bf{m}},6}^z
\right.
\nonumber\\
\left.
+\frac{J_{26}-J_{36}}{2}\left(T^x_{\bf{m}}s_{{\bf{m}},6}^x+T^y_{\bf{m}}s_{{\bf{m}},6}^y\right)
\right.
\nonumber\\
\left.
+\frac{J_{61}+J_{64}-2J}{4}s_{{\bf{m}},6}^zT^z_{m_x,m_y+1}
\right.
\nonumber\\
\left.
+\frac{-J_{61}+J_{64}}{2}\left(s_{{\bf{m}},6}^xT^x_{m_x,m_y+1}+s_{{\bf{m}},6}^yT^y_{m_x,m_y+1}\right)
\right\}
\end{eqnarray}
(i.e., ${\cal{V}}={\cal{H}}-{\cal{H}}_{\rm{main}}$).
The ground state $\vert\varphi_0\rangle$ of the Hamiltonian ${\cal{H}}_{\rm{main}}$ 
(then $s_{{\bf{m}},5}^z=s_{{\bf{m}},6}^z=1/2$)
has the energy $\varepsilon_0=-(5J_2+J){\cal{N}}$.
It is $2^{{\cal{N}}}$-fold degenerate
(since it does not depend on the value of $T^z_{\bf{m}}=\pm 1/2$) 
and forms a model space defined by the projector $P=\vert\varphi_0\rangle\langle\varphi_0\vert$.
Explicitly this projector can be written as
\begin{eqnarray}
\label{207}
P=\otimes_{{\bf{m}}}P_{{\bf{m}}},
\nonumber\\
P_{\bf{m}}={\cal{P}}_{\bf{m}}\otimes \left(\vert\uparrow_5\rangle\langle\uparrow_5\vert \otimes \vert\uparrow_6\rangle\langle\uparrow_6\vert\right)_{\bf{m}}.
\end{eqnarray}
For $J_{15}-J \ne 0$, \ldots, $J_{64}-J \ne 0$,  and $h-h_1 \ne 0$ we construct an effective Hamiltonian ${\cal{H}}_{\rm{eff}}$ 
which acts on the model space only but which gives the exact ground-state energy. 
${\cal{H}}_{\rm{eff}}$ can be found perturbatively and it is given by\cite{klein,fulde,essler}
\begin{eqnarray}
\label{208}
{\cal{H}}_{\rm{eff}}
=P{\cal{H}}P+P{\cal{V}}\sum_{\alpha\ne 0}\frac{\vert \varphi_{\alpha}\rangle\langle \varphi_{\alpha}\vert}{\varepsilon_0-\varepsilon_{\alpha}}{\cal{V}}P+\ldots .
\end{eqnarray}
Here $\vert \varphi_{\alpha}\rangle$ ($\alpha\ne 0$) are the known excited states of ${\cal{H}}_{\rm{main}}$ (\ref{205}).
The set of relevant excited states which enters the second term in Eq.~(\ref{208})
is constituted of the states with one flipped spin on those sites that connect two neighboring squares.
The energy of the excited states depends on the states of these two squares.
Namely, it acquires 
the value $\varepsilon_\alpha=\varepsilon_0+2J_2-J$ 
if both squares are in the $\vert u\rangle$ state,
the value $\varepsilon_\alpha=\varepsilon_0+2J_2-J/2$ 
if one of the squares is in the $\vert u\rangle$ state and the other one in the $\vert d\rangle$ state,
and 
the value $\varepsilon_\alpha=\varepsilon_0+2J_2$ 
if both squares are in the $\vert d\rangle$ state.
Taking this into account, 
we can calculate the second term of Eq.~(\ref{208}) 
and after using (pseudo)spin operators (\ref{203}) we finally arrive at the Hamiltonian
\begin{eqnarray}
\label{209}
{\cal{H}}_{{\rm{eff}}}=\sum_{(mn)}
\left[{\sf{J}}_{mn}\left(T_m^xT_{n}^x + T_m^yT_{n}^y\right) +{\sf{J}}^z_{mn}T_m^zT_{n}^z\right]
\nonumber\\
-{\sf{h}}\sum_{m=1}^{{\cal{N}}}T^z_m +{\cal{N}}{\sf{C}},
\end{eqnarray}
where the first sum runs over the neighboring sites of an ${\cal{N}}$-site square lattice.
The parameters ${\sf{J}}_{mn}$, ${\sf{J}}^z_{mn}$, ${\sf{h}}$, and  ${\sf{C}}$ are given by
\begin{eqnarray}
\label{210}
{\sf{J}}_h=-\frac{\left(-J_{15}+J_{25}\right)\left(J_{54}-J_{53}\right)}{16J_2}\frac{1}{1-\frac{J}{2J_2}},
\nonumber\\
{\sf{J}}_v=-\frac{\left(J_{26}-J_{36}\right)\left(-J_{61}+J_{64}\right)}{16J_2}\frac{1}{1-\frac{J}{2J_2}},
\nonumber\\
{\sf{J}}^z_h=\frac{S_h}{16J_2}\left(\frac{1}{1-\frac{J}{2J_2}}-\frac{1}{1-\frac{J}{4J_2}}\right),
\nonumber\\
{\sf{J}}^z_v=\frac{S_v}{16J_2}\left(\frac{1}{1-\frac{J}{2J_2}}-\frac{1}{1-\frac{J}{4J_2}}\right),
\nonumber\\
{\sf{h}}=h-h_1-\frac{S_h+S_v}{16J_2}\frac{1}{1-\frac{J}{4J_2}},
\nonumber\\
{\sf{C}}=-\frac{5}{2}h+\frac{3}{2}J
\nonumber\\
-\frac{S_h+S_v}{64J_2}\left(\frac{1}{1-\frac{J}{2J_2}}+\frac{1}{1-\frac{J}{4J_2}}\right),
\nonumber\\
J=\frac{J_{15}+J_{25}+J_{54}+J_{53}+J_{26}+J_{36}+J_{61}+J_{64}}{8},
\nonumber\\
S_h=\frac{\left(-J_{15}+J_{25}\right)^2+\left(J_{54}-J_{53}\right)^2}{2},
\nonumber\\
S_v=\frac{\left(J_{26}-J_{36}\right)^2+\left(-J_{61}+J_{64}\right)^2}{2},
\nonumber\\
h_1=2J_2+J.
\end{eqnarray}
Here the index $h$ ($v$) corresponds to the horizontal (vertical) direction.
For the special case
$J_{25}=J_{54}=J_{36}=J_{61}=J_1$ and $J_{15}=J_{53}=J_{26}=J_{64}=J_3$
Eqs.~(\ref{209}) and (\ref{210}) transform into Eqs.~(3.10) and (3.11) of Ref.~\onlinecite{drkk}.

In the limit $J/J_2 \to 0$ 
Eqs.~(\ref{209}) and (\ref{210}) transform into the effective Hamiltonian $H_{\rm{eff}}$ obtained by the strong-coupling approximation 
with the parameters 
\begin{eqnarray}
\label{211}
{\sf{J}}_h=-\frac{\left(-J_{15}+J_{25}\right)\left(J_{54}-J_{53}\right)}{16J_2},
\nonumber\\
{\sf{J}}_v=-\frac{\left(J_{26}-J_{36}\right)\left(-J_{61}+J_{64}\right)}{16J_2},
\nonumber\\
{\sf{J}}^z_h={\sf{J}}^z_v=0,
\nonumber\\
{\sf{h}}=h-h_1-\frac{S_h+S_v}{16J_2},
\nonumber\\
{\sf{C}}=-\frac{5}{2}h+\frac{3}{2}J-\frac{S_h+S_v}{32J_2}.
\end{eqnarray}
In this limit  the effective Hamiltonian is the square-lattice spin-1/2 isotropic $XY$ model in a transverse magnetic field. 
Again Eq.~(\ref{211}) transforms into Eq.~(A7) of Ref.~\onlinecite{drkk} for the azurite-like nonideal geometry.

\begin{figure}
\begin{center}
\includegraphics[clip=on,width=85mm,angle=0]{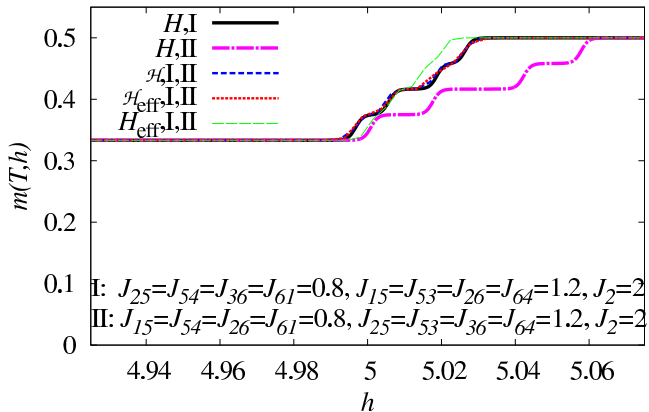}
\caption
{(Color online) 
Field dependence of the low-temperature magnetization per site $m$
of the full and effective models for the distorted square-kagome lattice of ${\cal{N}}=4$ cells ($T=0.001$).
The first set of parameters (I) corresponds to the azurite-like nonideal geometry:
$J_{25}=J_{54}=J_{36}=J_{61}=0.8$, 
$J_{15}=J_{53}=J_{26}=J_{64}=1.2$, 
$J_2=2$ [cf. Fig.~5(a) of Ref.~\onlinecite{drkk}].
The second set of parameters (II) is as follows:
$J_{15}=J_{54}=J_{26}=J_{61}=0.8$, 
$J_{25}=J_{53}=J_{36}=J_{64}=1.2$, 
$J_2=2$.}
\label{fig02}
\end{center}
\end{figure}

The considered case of a general nonideal geometry allows us to discuss the quality of the elaborated effective description.
The effective theories are based on accounting of only two states for each square, $\vert u\rangle$ and $\vert d\rangle$,
and may overestimate a tendency for localization.
This can be seen already from inspection of the constants ${\sf{J}}_h$ and ${\sf{J}}_v$ given in Eqs.~(\ref{210}) and (\ref{211}).
According to these formulas, 
having, for example, $J_{15}=J_{25}$ and $J_{26}=J_{36}$ but $J_{53}\ne J_{54}$ and $J_{61}\ne J_{64}$
would be sufficient to suppress completely a spreading of localized states over the lattice.
By contrast, exact-diagonalization results (not shown here) demonstrate  that this condition
is not sufficient to avoid the spreading, 
rather we need in addition the equalities $J_{53}=J_{54}$ and $J_{61}=J_{64}$.
In Fig.~\ref{fig02} we compare exact-diagonalization data
 for the low-temperature magnetization curve
calculated for the full model of $N=24$ sites
for two sets of parameters,
$J_{25}=J_{54}=J_{36}=J_{61}=0.8$, $J_{15}=J_{53}=J_{26}=J_{64}=1.2$, $J_2=2$ 
[the azurite-like nonideal geometry, cf. Fig.~5(a) of Ref.~\onlinecite{drkk}]
and
$J_{15}=J_{54}=J_{26}=J_{61}=0.8$, $J_{25}=J_{53}=J_{36}=J_{64}=1.2$,
$J_2=2$, with the predictions obtained from corresponding effective models (\ref{204}), (\ref{209}), (\ref{210}), and (\ref{209}), (\ref{211}). 
For this choice both sets yield identical results within each of the
effective theories, but both sets lead to different results  for the full
initial model.
While for the first set the effective models (except the strong
coupling-approximation) work well, see the short-dashed blue curves and dotted red
curves in Fig.~\ref{fig02}, for the second data set the agreement is less
satisfactorily, since the initial model exhibits a wider field region where magnetization varies between the two plateau values, $m=1/3$ and $m=1/2$
(dash-dotted magenta curve in Fig.~\ref{fig02}).
In the latter case a discrepancy emerges already between the results which follow from $H$ (\ref{201}) and ${\cal{H}}$
(\ref{202}), which leads to the conclusion 
that the restriction to only two states of each square yields excellent or only modest
results depending on specific nonideal geometry under consideration.

Similar to model (\ref{204}), 
the obtained spin lattice models (\ref{209}), (\ref{210}) and (\ref{209}), (\ref{211}) are also unfrustrated, 
however, they are simpler and have less sites ${\cal{N}}=N/6$.
Therefore they are more appropriate for further analysis using, for example, quantum Monte Carlo techniques.
We will report such  quantum Monte Carlo results in the next section.

\section{High-field low-temperature specific heat and phase diagram}
\label{sec3}
\setcounter{equation}{0}

After having derived the effective models (\ref{209}), (\ref{210}) and (\ref{209}), (\ref{211}),
the beautiful results known for the square-lattice spin-1/2 $XX0$/$XXZ$ Heisenberg model in a $z$-aligned magnetic field
can be used to understand the high-field low-temperature properties of the square-kagome quantum Heisenberg antiferromagnet. 
Since we know that our effective models work very well for azurite-like distortions of ideal geometry   
(see the above discussion),  
in what follows we restrict ourselves to this case 
and consider the specific parameter set $J_1=0.8$, $J_2=2$, and $J_3=1.2$.
For this set of parameters exact-diagonalization data have been reported already in Ref.~\onlinecite{drkk}.
However, those results were restricted to small systems up to ${\cal{N}}=20$.
Now we again consider the effective model ${\cal{H}}_{\rm{eff}}$ given in Eqs.~(\ref{209}) and (\ref{210})
but present results of more time-demanding quantum Monte Carlo calculations 
which refer to much larger systems up to ${\cal{N}}=24\times24=576$.
To perform these calculations we used the {\tt{dirloop\underline{\space}sse}} package 
(the directed loop algorithm in the stochastic series expansion representation) 
from the ALPS library.\cite{alps}
For concreteness we will focus on the temperature dependence of the specific heat per site $c(T,h)$
which may be quite sensible to the system size ${\cal{N}}$
(in contrast, for example, to the temperature dependence of the magnetization which shows no size effect).
It should be noted that the effective spin models were investigated via quantum Monte Carlo simulations in the past for particular parameter sets. 
Note further that these models are also considered in the context of hard-core bosons, 
where the $z$-aligned magnetization corresponds to the particle number and the magnetic field to the chemical potential.\cite{note1}
Although these previous studies\cite{qmc_d0,qmc_d1,qmc_d2,harada,qmc_s,tognetti,hcbosons_1,hcbosons_2,rigol,vicari} provide a physical picture in general,
we have here to perform specific calculations for the effective model with special parameter sets 
corresponding to the distorted square-kagome quantum Heisenberg antiferromagnet at hand. 

\begin{figure}
\begin{center}
\includegraphics[clip=on,width=85mm,angle=0]{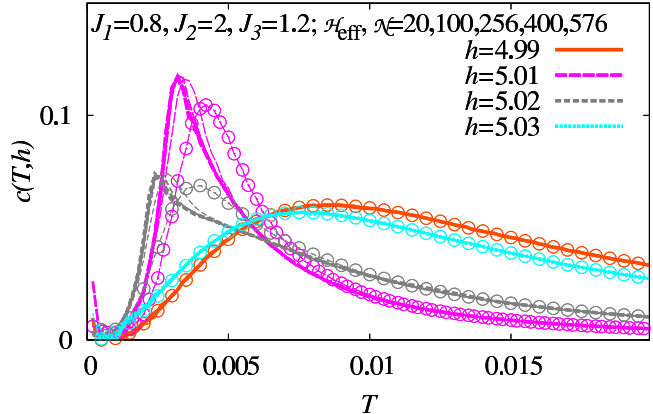}
\caption
{(Color online) 
Specific heat per site $c(T,h)$ at high fields ($h=4.99,5.01,5.02,5.03$) and low temperatures
for the distorted square-kagome Heisenberg antiferromagnet (\ref{201}) with $J_1=0.8$, $J_2=2$, $J_3=1.2$
obtained by  quantum Monte Carlo simulations for the effective Hamiltonian ${\cal{H}}_{\rm{eff}}$ (\ref{209}), (\ref{210}) 
with ${\cal{N}}=100,256,400,576$ (the line thickness increases with increase of ${\cal{N}}$).
For comparison, we also show  exact-diagonalization data for ${\cal{N}}=20$
by very thin curves with circles
[cf. Fig.~11(b) of Ref.~\onlinecite{drkk}].}
\label{fig03}
\end{center}
\end{figure}

From preceding studies (see Ref.~\onlinecite{drkk})
we know that deviations from the ideal geometry lead to the following modifications
in a small region of $h$ around the saturation field  $h_{\rm{sat}}=h_1$.
(i) Instead of the jump of the ground-state magnetization $m(T=0)$ at $h_1$ from a plateau at $m=1/3$ to saturated magnetization  $m=1/2$,
there is a small finite region $h_{1l}\le h\le h_{1h}$ around $h_1$
(for $J_1=0.8$, $J_2=2$, and $J_3=1.2$ we have $h_1=5$, $h_{1l}\approx 4.996$, $h_{1h}\approx 5.027$)
where the magnetization shows a steep increase between the two plateau values, $m=1/3$ and $m=1/2$.
(ii) Instead of a nonzero residual entropy at $h_1$,
there is zero residual entropy followed by a strong enhancement of the entropy at very small temperatures.
(iii) Instead of zero specific heat at $h_1$,
the specific heat $c(T)$ shows a $T^2$-decay as $T\to 0$ in a small region of $h$ around $h_1$, 
but it vanishes exponentially as $T\to 0$ in the plateau regions, i.e., in the gapped phase.
Our quantum Monte Carlo results (${\cal{N}}=100,\,256,\,400,\,576$) for the specific heat collected in Fig.~\ref{fig03}
support this scenario for the low-temperature behavior of $c(T,h)$.
In particular, from Fig.~\ref{fig03} one can find indications for different decay laws as $T\to 0$ 
for $h=4.99$ and $h=5.03$ (exponential) 
and
for $h=5.01$ and $h=5.02$ (power-law).
Note, however, that at extremely low temperatures quantum Monte Carlo data become noisy 
that restricts our consideration to temperatures above $T=0.0005$.
(We note that exact-diagonalization data at extremely low temperatures become also unreliable, since they suffer from finite-size artifacts.)
New features with respect to our previous exact-diagonalization study\cite{drkk} appearing for larger systems 
are obvious from Fig.~\ref{fig03}, 
where for comparison also  exact-diagonalization data for ${\cal{N}}=20$ are shown. 
As ${\cal{N}}$ increases, the peak for $h=5.01$ and $h=5.02$  becomes somewhat higher and sharper and moves to slightly lower temperatures.
For $h=5.02$ it changes even its form.
On the other hand, 
the temperature profiles for $h=4.99$ and $h=5.03$ are insensitive to the system sizes.
This behavior of temperature profiles as $h$ varies reflects the difference in the low-temperature specific heat 
for 
the gapless phase ($h$ is inside the region $h_{1l}\le h\le h_{1h}$) 
and 
the gapped phases ($h$ is outside this region).

The most intriguing property of the effective models (\ref{209}), (\ref{211}) and (\ref{209}), (\ref{210}) 
is the existence of a BKT transition.
A classical two-dimensional isotropic $XY$ model undergoes a transition 
from bound vortex-antivortex pairs at low temperatures to unpaired vortices and antivortices at some critical temperature $T_c$.\cite{bkt}
For $T<T_c$ 
(superfluid phase) 
the system is characterized by quasi-long-range order,
i.e., correlations decay algebraically at large distances without the emergence of a nonvanishing order parameter.
For $T>T_c$ 
(normal phase)
the system is disordered with an exponential increase of the correlation length $\xi$ as $T\to T_c$,
\begin{eqnarray}
\label{301}
\xi\propto  e^{\frac{b}{\sqrt{\tau}}},
\;\;\;
\tau=\frac{T-T_c}{T_c}.
\end{eqnarray}
The BKT transition temperature for the classical square-lattice isotropic $XY$ model (without field) 
is $T_c\approx 0.893 \vert{\sf{J}}\vert$.\cite{gupta,hasenbusch,komura,hsieh}
Within numerical studies dealing with finite systems 
it is quite difficult to extract an exponential divergence of $\xi$ at $T_c$ 
from the finite-size data for the large-distance behavior of spin correlations.
Another important quantity to pin down the BKT transition
is the so-called helicity modulus $\Upsilon$ which is related to the superfluid density $\rho_s$.\cite{fisher}
In the quantum spin-1/2 case, the BKT critical behavior occurs too.\cite{qmc_d0,qmc_d1,harada}
The critical temperature for the $s=1/2$ case is estimated as $T_c\approx 0.34 \vert{\sf{J}}\vert$.\cite{harada,rigol,vicari}
The quantum model is gapless with an excitation spectrum that is linear in the momentum.
The specific heat $c(T)$ shows $T^2$ behavior for $T \to 0$, 
it increases very rapidly around $T_c$, 
and it exhibits a finite peak somewhat above $T_c$.\cite{qmc_d1}
This kind of the low-temperature thermodynamics survives for not too large $z$-aligned magnetic field $\vert {\sf{h}}\vert < 2\vert {\sf{J}}\vert$
(see,  e.g., Figs.~3 and 8 in Ref.~\onlinecite{hcbosons_2} or Fig.~1 in Refs.~\onlinecite{rigol} and \onlinecite{vicari}).
Also for the spin-1/2 square-lattice $XXZ$ model with dominating isotropic $XY$ interaction in a $z$-aligned magnetic field the BKT transition appears.\cite{hcbosons_1}

\begin{figure}
\begin{center}
\includegraphics[clip=on,width=85mm,angle=0]{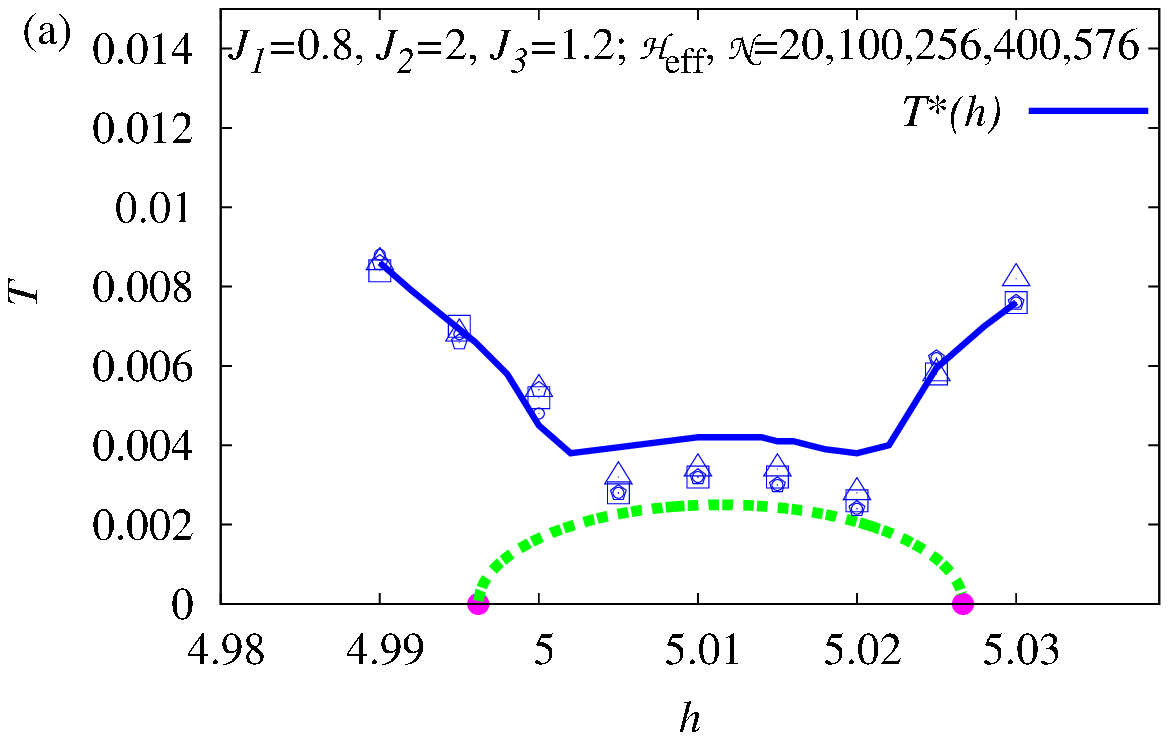}\\
\vspace{5mm}
\includegraphics[clip=on,width=85mm,angle=0]{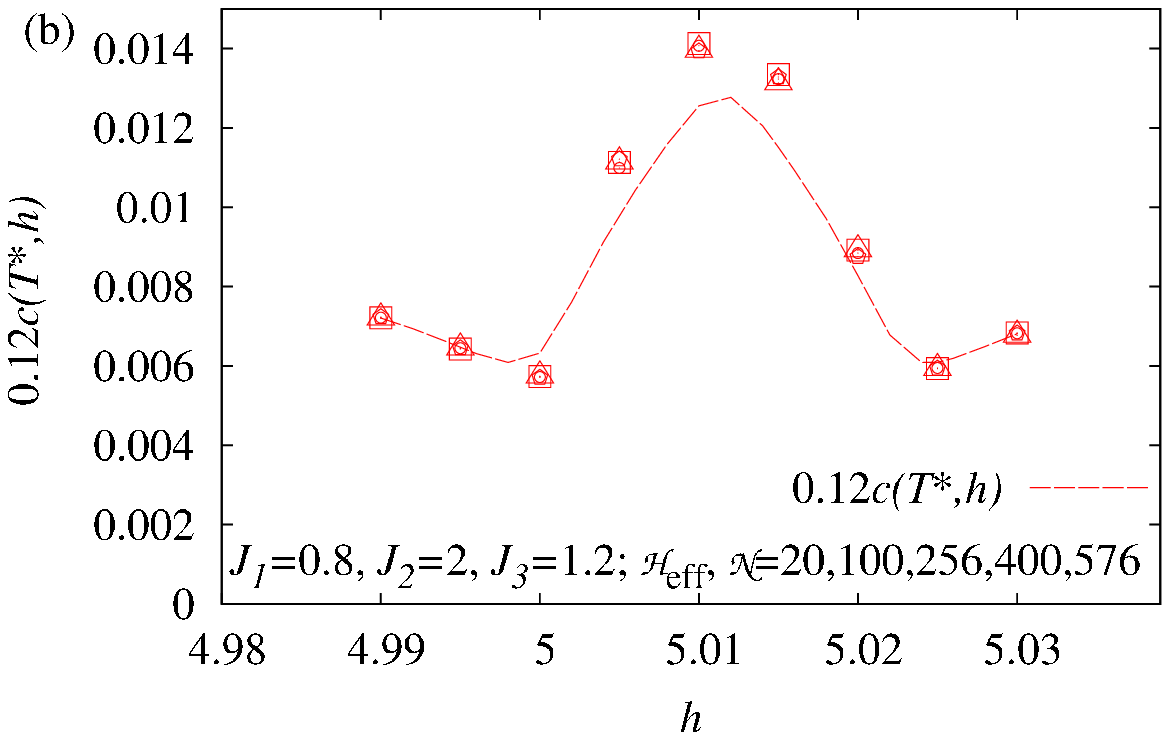}\\
\vspace{5mm}
\includegraphics[clip=on,width=85mm,angle=0]{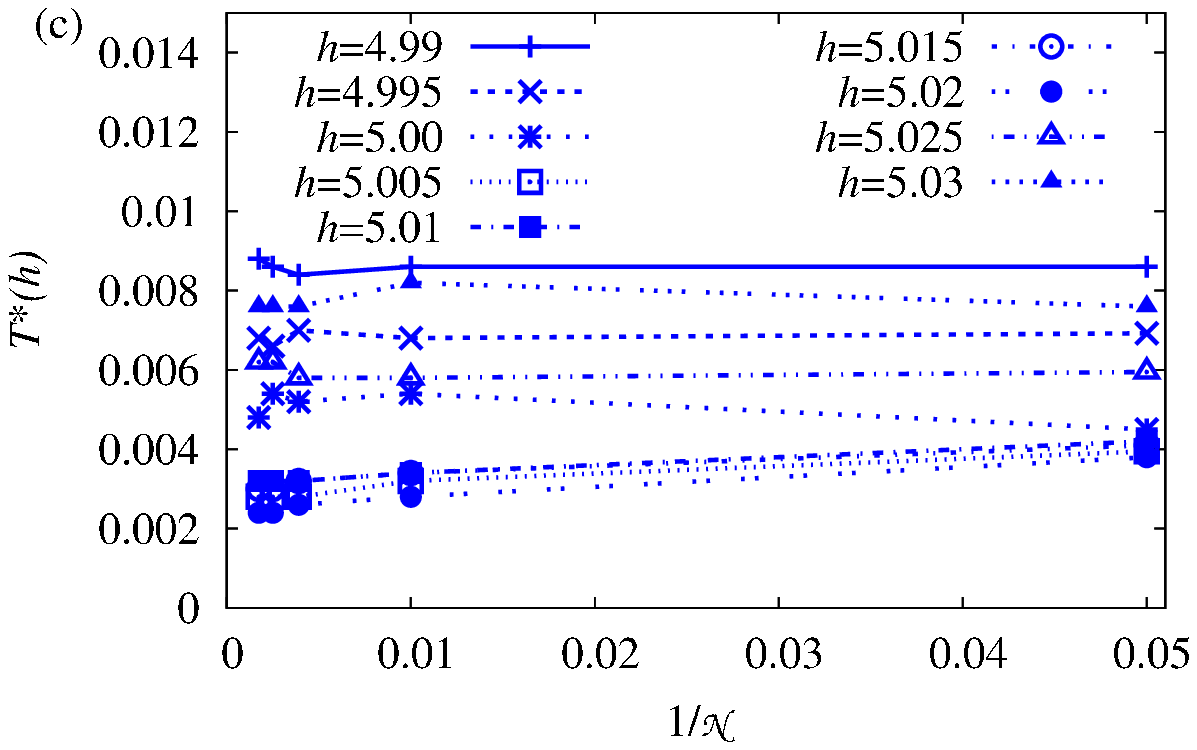}
\caption
{(Color online) 
(a)
Sketch of the phase diagram of the distorted square-kagome Heisenberg antiferromagnet 
($J_1=0.8$, $J_2=2$, $J_3=1.2$) 
at high magnetic field  (thick dashed green line)
as it is indicated by the position of the maximum $T^{*}$ of the specific heat $c(T,h)$ 
shown by the thick solid blue curve (${\cal{N}}=20$) 
and the blue symbols 
(triangles -- ${\cal{N}}=100$, 
squares -- ${\cal{N}}=256$, 
pentagons -- ${\cal{N}}=400$, 
circles -- ${\cal{N}}=576$).
By filled violet circles the values of $h_{1l}\approx 4.996$ and $h_{1h}\approx 5.027$ are indicated.
(b)
Height of the maximum in the specific heat $c(T^*,h)$ (multiplied by 0.12): 
dashed red curve -- ${\cal{N}}=20$,
red triangles -- ${\cal{N}}=100$, 
squares -- ${\cal{N}}=256$, 
pentagons -- ${\cal{N}}=400$, 
circles -- ${\cal{N}}=576$.
(c) 
Dependence of $T^{*}(h)$ on $1/{\cal{N}}$ at various fields $h$.}
\label{fig04}
\end{center}
\end{figure}

Following our previous study,\cite{drkk}
we use the observation of Ref.~\onlinecite{qmc_d1} 
that the BKT transition point $T_c$ is located somewhat below the well-pronounced peak-like maximum of the specific heat.
Although the adopted criterion to fix the critical temperature $T_c$ for different $h$ is a rather rough one,
it can provide a sketch of the phase diagram based on specific-heat data.
Since the peak in $c(T)$ calculated by exact diagonalization shows noticeable finite-size effects,\cite{drkk}
we use here the quantum Monte Carlo approach to obtain  data for much larger systems 
thus getting more accurate predictions.
A sketch of the phase diagram of the distorted square-kagome Heisenberg antiferromagnet in the  $h$ -- $T$ plane
which uses the maximum in the specific heat as an indicator of the BKT transition is reported in Fig.~\ref{fig04}(a). 
In this figure 
the thick solid blue line
corresponds to the position $T^*$ of the maximum in the specific heat per site $c(T,h)$
obtained by exact diagonalization earlier for ${\cal{N}}=20$
(cf. Fig.~12 of Ref.~\onlinecite{drkk}).
The blue symbols correspond to quantum Monte Carlo data (${\cal{N}}=100,\,256,\,400,\,576$) for $T^*(h)$. 
Based on these data for $T^*$ we have drawn  
the thick dashed green line showing a tentative BKT-transition line $T_c(h)$.
Fig.~\ref{fig04}(b) shows the  height of the maximum  in the specific heat, 
i.e.,    
the value of $c(T^*,h)$ (here multiplied by 0.12 to get correspondence to Fig.~12 of Ref.~\onlinecite{drkk}).
Again we compare exact-diagonalization data for ${\cal{N}}=20$ (thin red
line) with new quantum Monte Carlo data for ${\cal{N}}=100,\,256,\,400,$ and $576$ (symbols). 
It is obvious that the height of the maximum increases
noticeably in the field region where a BKT transition appears.  
Finally, we illustrate the finite-size dependence of $T^*(h)$ in
Fig.~\ref{fig04}(c), which is obviously weak.     
Although $T^{*}(h)$, $h_{1l} \le h \le h_{1h}$, 
is shifted to slightly lower temperatures for large values ${\cal{N}}$
in comparison with the previous prediction,\cite{drkk}
the values of $T^*(h)$ apparently are already close to their values in the limit
${\cal{N}}\to\infty$.

\section{Conclusions}
\label{sec4}
\setcounter{equation}{0}

In the present paper we have improved 
the low-energy theory of the almost flat-band square-kagome quantum Heisenberg antiferromagnet at high magnetic fields.
Analytical results for effective Hamiltonians refer now to (small) deviations
of general case from the flat-band situation.
The relevant effective model has been investigated for quite large system sizes using quantum Monte Carlo simulations.
The high-field low-temperature phase diagram of the distorted square-kagome quantum Heisenberg antiferromagnet reported in Fig.~\ref{fig04}(a)  
refines previous findings which were based on exact-diagonalization data for small systems.
Although the existence of the BKT transition is not questionable, the precise phase diagram remains an open question.
To find precise values for the BKT-transition temperature for the square-kagome quantum Heisenberg antiferromagnet, 
in fact, one has to determine accurately, e.g., by quantum Monte Carlo techniques, 
the BKT-transition temperature of the corresponding effective square-lattice spin-1/2 $XXZ$ easy-plane model 
in a $z$-aligned magnetic field.  
To the best of our knowledge, 
this problem has not been studied yet and its consideration is out of the scope of this article.
Finally,
the reported results shed more light on possible manifestation of localized-magnon effects in experiments,
if a realization of the square-kagome Heisenberg antiferromagnet becomes available.

\section*{Acknowledgments}

The present study was supported by the DFG (project RI615/21-1).
O.~D. acknowledges the kind hospitality of the University of Magdeburg in October-December of 2012 and in March-May of 2013.
O.~D. acknowledges financial support 
of the organizers of the 15$^{\rm{th}}$ CSMAG conference (Ko\v{s}ice, 17-21 June 2013) 
and of the Pavol Jozef \v{S}af\'{a}ric University in Ko\v{s}ice,
and of the Abdus Salam International Centre for Theoretical Physics (Trieste, August, 2013).
O.~D. and J.~R. are grateful to the MPIPKS (Dresden) for the kind hospitality in October-December of 2013.

\appendix
\section{Effective Hamiltonians for the quantum Heisenberg antiferromagnet 
         on a distorted frustrated diamond chain at high magnetic fields}
\label{a}
\renewcommand{\theequation}{\thesection.\arabic{equation}}
\setcounter{equation}{0}

\begin{figure}[ht]
\begin{center}
\includegraphics[clip=on,width=70mm,angle=0]{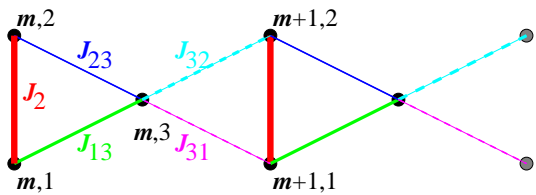}
\caption
{(Color online) 
The diamond chain described by Hamiltonian (\ref{201}).
The trapping cells for localized magnons (vertical dimers) are indicated by bold red lines ($J_2$ bonds).}
\label{fig06}
\end{center}
\end{figure}

In this appendix we provide a similar extension for the one-dimensional counterpart of the square-kagome lattice, 
namely the distorted diamond chain, for completeness and comparison.
In the case of the diamond chain\cite{diamond} with a most general exchange coupling scheme
[see Eq.~(\ref{201}) and Fig.~\ref{fig06}]
we arrive at the following results.
The effective Hamiltonian ${\cal{H}}$ [cf. Eq.~(\ref{204})] reads
\begin{eqnarray}
\label{a01}
{\cal{H}}
=\sum_{m=1}^{{\cal{N}}}
\left[ 
-\frac{h}{2}-\frac{J_2}{4}-\left(h-J_2\right)T^z_m -\left(h-J\right)s^z_{m,3} 
\right.
\nonumber\\
\left.
+\frac{J_{13}+J_{23}}{2}T^z_ms^z_{m,3}
\right.
\nonumber\\
\left.
+\frac{-J_{13}+J_{23}}{\sqrt{2}} \left(T^x_ms^x_{m,3}+T^y_ms^y_{m,3}\right)
\right.
\nonumber\\
\left.
+\frac{J_{31}+J_{32}}{2}s^z_{m,3}T^z_{m+1}
\right.
\nonumber\\
\left.
-\frac{J_{31}-J_{32}}{\sqrt{2}} \left(s^x_{m,3}T^x_{m+1}+s^y_{m,3}T^y_{m+1}\right)
\right],
\nonumber\\
J=\frac{J_{13}+J_{23}+J_{31}+J_{32}}{4} \;. \; \;
\end{eqnarray}
The effective Hamiltonian ${\cal{H}}_{{\rm{eff}}}$ [cf. Eqs.~(\ref{209}), (\ref{210})] is given by the formula
\begin{eqnarray}
\label{a02}
{\cal{H}}_{{\rm{eff}}}=\sum_{m=1}^{{\cal{N}}}
\left[{\sf{J}}\left(T_m^xT_{m+1}^x + T_m^yT_{m+1}^y\right) +{\sf{J}}^zT_m^zT_{m+1}^z
\right.
\nonumber\\
\left.
-{\sf{h}}T^z_m +{\sf{C}} \; \right ]\quad
\end{eqnarray}
with the following parameters
\begin{eqnarray}
\label{a03}
{\sf{J}}=\frac{\left(-J_{13}+J_{23}\right)\left(J_{31}-J_{32}\right)}{4J_2}\frac{1}{1-\frac{J}{J_2}},
\nonumber\\
{\sf{J}}^z=\frac{\left(-J_{13}+J_{23}\right)^2+\left(J_{31}-J_{32}\right)^2}{8J_2}
\left(\frac{1}{1-\frac{J}{J_2}}-1\right),
\nonumber\\
{\sf{h}}=h-h_1-\frac{\left(-J_{13}+J_{23}\right)^2+\left(J_{31}-J_{32}\right)^2}{8J_2},
\nonumber\\
{\sf{C}}=-h-\frac{J_2}{4}+\frac{J}{2}
\nonumber\\
-\frac{\left(-J_{13}+J_{23}\right)^2+\left(J_{31}-J_{32}\right)^2}{32J_2}
\left(\frac{1}{1-\frac{J}{J_2}}+1\right),
\nonumber\\
J=\frac{J_{13}+J_{23}+J_{31}+J_{32}}{4},
\nonumber\\
h_1=J_2+J.
\;
\end{eqnarray}
In the limit $J_{23}=J_{31}=J_1$ and $J_{13}=J_{32}=J_3$ these results coincide with those ones obtained in Ref.~\onlinecite{drkk},
see Eqs.~(3.6) and (3.7) of Ref.~\onlinecite{drkk}.

\begin{figure}
\begin{center}
\includegraphics[clip=on,width=85mm,angle=0]{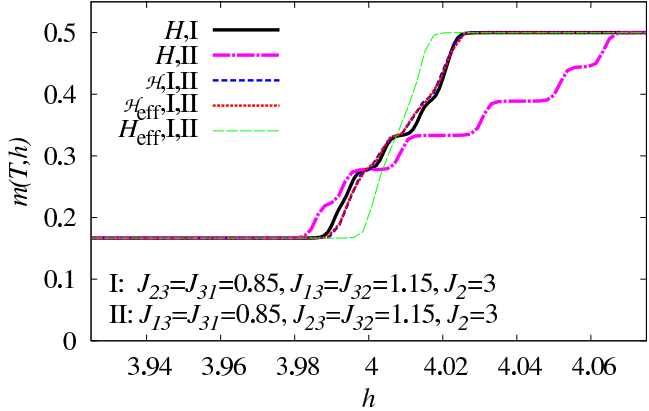}
\caption
{(Color online) 
Field dependence of the low-temperature magnetization per site
of the full and effective models for the distorted diamond chain of ${\cal{N}}=6$ cells ($T=0.001$).
The first set of parameters (I) corresponds to the azurite-like nonideal geometry:
$J_{23}=J_{31}=0.85$, 
$J_{13}=J_{32}=1.15$, 
$J_2=3$ [cf. Fig.~3(a) of Ref.~\onlinecite{drkk}].
The second set of parameters (II) is as follows:
$J_{13}=J_{31}=0.85$, 
$J_{23}=J_{32}=1.15$, 
$J_2=3$.}
\label{fig07}
\end{center}
\end{figure}

Similar to the case of the square-kagome lattice,
effective theories overestimate the tendency of localization.
For example, 
$J_{13}=J_{23}$ but $J_{31}\ne J_{32}$ or vice versa is sufficient to suppress completely a spreading of localized states within the effective models.
On the other hand, exact-diagonalization data for the full model
indicate that this condition
is not sufficient to suppress the spreading, rather we need both equalities to hold,
$J_{13}=J_{23}$ but $J_{31}=J_{32}$.
In Fig.~\ref{fig07} we compare exact-diagonalization data for the low-temperature magnetization curve 
for the initial full model and the effective models
considering two sets of parameters:
$J_{23}=J_{31}=0.85$, $J_{13}=J_{32}=1.15$, $J_2=3$ [cf. Fig.~3(a) of Ref.~\onlinecite{drkk}]
and
$J_{13}=J_{31}=0.85$, $J_{23}=J_{32}=1.15$, $J_2=3$.
Each effective model yields identical predictions for both sets of parameters,
whereas the results for the initial model are different
(compare solid black and dash-dotted magenta curves in Fig.~\ref{fig07}).
Furthermore,
for the azurite-like nonideal geometry the effective theory based on Eqs.~(\ref{a02}), (\ref{a03})
provides a quite good quantitative description of the magnetization curve,
whereas in the other case the agreement is only qualitative.

\end{document}